\documentclass[aps,prd,12pt,amssymb]{revtex4}
                                                                          
\begin{document}                                                                         
                                                                          
\baselineskip=0.60cm
                                                                          
\newcommand{\ini}{\begin{equation}}
\newcommand{\fin}{\end{equation}}
\newcommand{\inir}{\begin{eqnarray}}
\newcommand{\finr}{\end{eqnarray}}
                                                                                
\def\ol{\overline}
\def\pa{\partial}
\def\ra{\rightarrow}
\def\ts{\times}
\def\df{\dotfill}
\def\bs{\backslash}
\def\dg{\dagger}
\def\la{\lambda}
\def\ep{\epsilon}
                                                                                
$~$
      
\vspace{1 cm}
                                                                               
\title{Seesaw and leptogenesis: a triangular ansatz}

\author{D. Falcone}

\affiliation{Dipartimento di Scienze Fisiche,
Universit\`a di Napoli, Via Cintia, Napoli, Italy}
                                                                                
\begin{abstract}
\vspace{1cm}
\noindent
A triangular ansatz for the seesaw mechanism and baryogenesis via leptogenesis is explored.
In a basis where both the charged lepton and the Majorana mass matrix are diagonal, the Dirac
mass matrix can generally be written as the product of a unitary times a triangular matrix. We assume
the unitary matrix to be the identity and then an upper triangular Dirac matrix.
Constraints from bilarge lepton mixing and leptogenesis are studied.
\end{abstract}
                                                                                
\maketitle
                                                                                
\newpage

\section{Introduction}

The baryogenesis via leptogenesis \cite{fy,lu,co,ba} is a framework 
proposed to explain the baryon asymmetry of the universe
without grand unification, but works well even within the SO(10) model. It is so important because is a consequence of the seesaw mechanism
\cite{ss,ms}, which in turn can explain the smallness of neutrino mass 
with respect to charged fermions.

In fact, the existence of very heavy Majorana neutrinos can generate small neutrino masses through mixing with Dirac neutrinos with masses similar to charged fermions (seesaw), and can produce a lepton asymmetry by means of decays to leptons and scalars (leptogenesis).
Sphalerons convert roughly one half of this lepton asymmetry to a baryon asymmetry (baryogenesis).

In a previous paper \cite{bfo}, quark-lepton symmetry 
between Dirac mass matrices was considered together with diagonal and 
offdiagonal forms of the heavy Majorana mass matrix.

In the present paper, starting from a general formalism for mass matrices, we explore an ansatz leading to an upper triangular form for the Dirac neutrino mass matrix in a basis where both the charged lepton and the Majorana neutrino mass matrices are diagonal. In ref.\cite{bra}
a lower triangular Dirac matrix is studied. This case is less predictive and the authors insert a texture zero.

\section{The seesaw mechanism}

According to the seesaw mechanism, the effective mass matrix of neutrinos
is given by the formula
\ini
M_{\nu} \simeq M_D M_R^{-1} M_D^T,
\fin
where $M_D$ is the Dirac mass matrix and $M_R$ the Majorana mass matrix.
For $M_R \gg M_D$, we have $M_{\nu} \ll M_D$.

In the standard model $M_D$ is generated by the coupling with the same Higgs doublet
that produces the up-quark mass matrix $M_u$, so that we expect the overall scale of
$M_D$ to be similar to the one of $M_u$. On the other hand, $M_R$ is generated as a
bare mass term (or by coupling to a Higgs singlet), so that its overall scale can be much
larger than the weak scale.

\section{Leptogenesis}

We consider leptogenesis formulas in the single-flavor approximation 
(see \cite{abada}). 
The baryon asymmetry, baryon to entropy fraction, is given by
\ini
Y_B \simeq \frac{1}{2} Y_L
\fin
and the lepton asymmetry by
\ini
Y_L \simeq 0.3 ~\frac{\epsilon_1}{g_*}
\left( \frac{0.55 \cdot 10^{-3} eV}{\tilde{m}_1} \right)^{1.16}
\fin
in the strong washout regime, or
\ini
Y_L \simeq 0.3 ~\frac{\epsilon_1}{g_*}
\left( \frac{\tilde{m}_1}{3.3 \cdot 10^{-3}eV} \right)
\fin
in the opposite weak washout regime. The parameter  $g_*$ is the number of light degrees of freedom, of the order $g_* \simeq 100$ in the standard case. 
Strong washout is realized for $\tilde{m}_1 \gg 3 \cdot 10^{-3}$,
where $\tilde{m}_1 = (M_D^{\dg} M_D)_{11} /M_1$. \par Notice that $Y_B$ is smaller than the baryon to photon ratio $\eta$ by roughly a factor 7. The experimental value of the baryon asymmetry is (see \cite{wmap}),
\ini
(Y_B)_{exp} \simeq 9 \cdot 10^{-11}.
\fin 
The CP-violating asymmetry $\epsilon_1$, related to the decay of the lightest right-handed neutrino, is given by
\ini
\epsilon_1 \simeq \frac{3}{16 \pi v^2} \left(
\frac{\text{Im} (M_D^{\dg}M_D)^2_{12}}{(M_D^{\dg}M_D)_{11}}
\frac{M_1}{M_2}+
\frac{\text{Im} (M_D^{\dg}M_D)^2_{13}}{(M_D^{\dg}M_D)_{11}}
\frac{M_1}{M_3} \right),
\fin
in the case $M_1 < M_2 \ll M_3$.

\section{The triangular model}

We take a basis where both the charged lepton and the right-handed neutrino mass matrices are diagonal:
\ini
M_e =
\left( \begin{array}{ccc}
m_e & 0 & 0 \\
0 & m_{\mu} & 0 \\
0 & 0 & m_{\tau}
\end{array} \right),
\fin

\ini
M_R =
\left( \begin{array}{ccc}
M_1 & 0 & 0 \\
0 & M_2 & 0 \\
0 & 0 & M_3
\end{array} \right).
\fin

The Dirac neutrino mass matrix can be written in the form $M_D = U Y_{\Delta} v$,
where $v=175$ GeV is the v.e.v. of the Higgs doublet, $U$ is a unitary matrix and $Y_{\Delta}$
is a triangular matrix
\ini
Y_{\Delta} =
\left( \begin{array}{ccc}
y_{11} & y_{12}  & y_{13}  \\
0 & y_{22}  & y_{23}  \\
0 & 0 & y_{33}
\end{array} \right),
\fin
with real diagonal elements and complex mixing elements.
The matrix $U$ cancels out in unflavored leptogenesis but not in the 
seesaw formula.

Our fundamental ansatz consists in taking $U=1$. Then the effective 
neutrino mass matrix becomes
\ini
M_{\nu} \simeq v^2
\left( \begin{array}{ccc}
\frac{y_{11}^2}{M_1}+\frac{y_{12}^2}{M_2}+\frac{y_{13}^2}{M_3} &
\frac{y_{12} y_{22} }{M_2}+\frac{y_{13} y_{23}}{M_3} &
\frac{y_{13} y_{33}}{M_3} \\
{*} & \frac{y_{22}^2}{M_2}+\frac{y_{23}^2}{M_3} & \frac{y_{23} y_{33}}{M_3} \\
{*} & {*} & \frac{y_{33}^2}{M_3}
\end{array} \right).
\fin
Parameters $y$ are determined imposing the phenomenological structure of 
the effective matrix  \cite{v,mr}
\ini
M_{\nu} \simeq v_0
\left( \begin{array}{ccc}
\epsilon^2 & \epsilon & \epsilon \\
\epsilon & 1 & 1 \\
\epsilon & 1 & 1
\end{array} \right),
\fin
with $\epsilon \simeq 0.05$, and $v_0 \simeq 0.02$.
Assuming the largest $y$ element to be of order 1, so that $v_0 \simeq v^2/M_3$,
we obtain
\ini
y_{33} \sim 1,~ y_{23} \sim 1,~ y_{13} \sim \epsilon
\fin
and
\ini
y_{22} \sim \sqrt({M_2/M_3}),~
y_{12} \sim \epsilon \sqrt({M_2/M_3}),~
y_{11} \sim \epsilon \sqrt({M_1/M_3}).
\fin
The Dirac mass matrix becomes
\ini
M_D \simeq v
\left( \begin{array}{ccc}
\epsilon \sqrt({M_1/M_3}) & \epsilon \sqrt({M_2/M_3})  & \epsilon  \\
0 & \sqrt({M_2/M_3}) & 1  \\
0 & 0 & 1
\end{array} \right).
\fin
It is useful to compare this matrix with the up-quark matrix in a basis where it is triangular and the down-quark matrix is diagonal \cite{kmw},
\ini
M_u \simeq m_t
\left( \begin{array}{ccc}
m_u/m_t & (m_c/m_t) V_{cd}  & V_{td}  \\
0 & m_c/m_t & V_{ts} \\
0 & 0 & 1
\end{array} \right),
\fin
showing deviation from quark-lepton symmetry in mass matrices.

We can now attempt to calculate the baryon asymmetry. First, we obtain the CP-asymmetry
\ini
\epsilon_1 = (3/16 \pi) (\epsilon^2) (M_1/M_3) (\sin \alpha + \sin \gamma),
\fin
where $\alpha$ is the phase angle of $y_{12}$ and $\gamma$ the phase angle of $y_{13}$.
Then, $\tilde{m}_1 \simeq 10^{-4}$, in the weak washout case. Finally, we get
\ini
Y_B \simeq 5 \cdot 10^{-5} ~\epsilon_1 \simeq 1.5 \cdot 10^{-4} ~(M_1/M_3) (\sin \alpha + \sin \gamma).
\fin
Matching with the value in (5), we obtain the relation
\ini
(M_1/M_3) (\sin \alpha + \sin \gamma) \simeq 6 \cdot 10^{-7}
\fin
and for $M_3 \simeq 10^{15}$ GeV also $(M_1) (\sin \alpha + \sin \gamma) \simeq 6 \cdot 10^{8}$ GeV.

\section{Conclusion}

Constraints from bilarge lepton mixing and leptogenesis determine a consistent relation
among masses and phases in a triangular ansatz of the seesaw mechanism.
Therefore, the rather strong condition $U=1$ is nevertheless viable.

$~$

We thank professors F. Buccella and L. Oliver for interesting discussions.

\end{document}